%% file: proc.tex
\documentclass[fleqn,12pt,twoside]{article}
\usepackage[headings]{espcrc1}

\readRCS
$Id: proc.tex,v 1.0 2005/09/20 11:22:11 alaszlo Exp $
\usepackage{graphicx}
\usepackage[figuresright]{rotating}

\newcommand{\AmS}{{\protect\the\textfont2
  A\kern-.1667em\lower.5ex\hbox{M}\kern-.125emS}}
\usepackage{blopeps}
\blopps{1mm}
\bloplw{0.25mm}

\title{High $p_T$ Spectra of Identified Particles Produced in 
       Pb+Pb Collisions at $158\,\mathrm{GeV/nucleon}$ Beam Energy}

\author{Andr\'as L\'aszl\'o and Tim Schuster 
        (for the \resizebox{0.9cm}{!}{\includegraphics{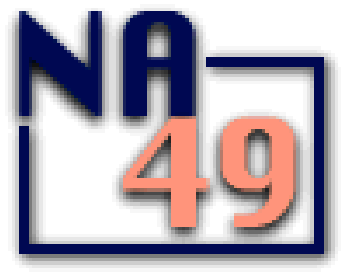}} 
        Collaboration)}
       
\runtitle{High $p_T$ Spectra of Identified Particles \dots}
\runauthor{A. L\'aszl\'o and T. Schuster}

\begin{document}

\maketitle
\vspace{0.5cm}
\noindent
C.~Alt$^{9}$, T.~Anticic$^{21}$, B.~Baatar$^{8}$,D.~Barna$^{4}$,
J.~Bartke$^{6}$, L.~Betev$^{10}$, H.~Bia{\l}\-kowska$^{19}$,
C.~Blume$^{9}$,  B.~Boimska$^{19}$, M.~Botje$^{1}$,
J.~Bracinik$^{3}$, R.~Bramm$^{9}$, P.~Bun\v{c}i\'{c}$^{10}$,
V.~Cerny$^{3}$, P.~Christakoglou$^{2}$, O.~Chvala$^{14}$,
J.G.~Cramer$^{16}$, P.~Csat\'{o}$^{4}$, P.~Dinkelaker$^{9}$,
V.~Eckardt$^{13}$, 
D.~Flierl$^{9}$, Z.~Fodor$^{4}$, P.~Foka$^{7}$,
V.~Friese$^{7}$, J.~G\'{a}l$^{4}$,
M.~Ga\'zdzicki$^{9,11}$, V.~Genchev$^{18}$, G.~Georgopoulos$^{2}$, 
E.~G{\l}adysz$^{6}$, K.~Grebieszkow$^{20}$,
S.~Hegyi$^{4}$, C.~H\"{o}hne$^{7}$, 
K.~Kadija$^{21}$, A.~Karev$^{13}$, M.~Kliemant$^{9}$, S.~Kniege$^{9}$,
V.I.~Kolesnikov$^{8}$, E.~Kornas$^{6}$, 
R.~Korus$^{11}$, M.~Kowalski$^{6}$, 
I.~Kraus$^{7}$, M.~Kreps$^{3}$, A.~Laszlo$^{4}$, M.~van~Leeuwen$^{1}$, 
P.~L\'{e}vai$^{4}$, L.~Litov$^{17}$, B.~Lungwitz$^{9}$,
M.~Makariev$^{17}$, A.I.~Malakhov$^{8}$, 
M.~Mateev$^{17}$, G.L.~Melkumov$^{8}$, A.~Mischke$^{1}$, M.~Mitrovski$^{9}$, 
J.~Moln\'{a}r$^{4}$, St.~Mr\'owczy\'nski$^{11}$, V.~Nicolic$^{21}$,
G.~P\'{a}lla$^{4}$, A.D.~Panagiotou$^{2}$, D.~Panayotov$^{17}$,
A.~Petridis$^{2}$, M.~Pikna$^{3}$, D.~Prindle$^{16}$,
F.~P\"{u}hlhofer$^{12}$, R.~Renfordt$^{9}$,
C.~Roland$^{5}$, G.~Roland$^{5}$, 
M. Rybczy\'nski$^{11}$, A.~Rybicki$^{6,10}$,
A.~Sandoval$^{7}$, N.~Schmitz$^{13}$, T.~Schuster$^{9}$, P.~Seyboth$^{13}$,
F.~Sikl\'{e}r$^{4}$, B.~Sitar$^{3}$, E.~Skrzypczak$^{20}$,
G.~Stefanek$^{11}$, R.~Stock$^{9}$, C.~Strabel$^{9}$, H.~Str\"{o}bele$^{9}$, T.~Susa$^{21}$,
I.~Szentp\'{e}tery$^{4}$, J.~Sziklai$^{4}$, P.~Szymanski$^{10,19}$,
V.~Trubnikov$^{19}$, D.~Varga$^{4,10}$, M.~Vassiliou$^{2}$,
G.I.~Veres$^{4,5}$, G.~Vesztergombi$^{4}$,
D.~Vrani\'{c}$^{7}$, A.~Wetzler$^{9}$,
Z.~W{\l}odarczyk$^{11}$ I.K.~Yoo$^{15}$, J.~Zim\'{a}nyi$^{4}$

\vspace{0.5cm}
\noindent
$^{1}$NIKHEF, Amsterdam, Netherlands. \\
$^{2}$Department of Physics, University of Athens, Athens, Greece.\\
$^{3}$Comenius University, Bratislava, Slovakia.\\
$^{4}$KFKI Research Institute for Particle and Nuclear Physics, Budapest, Hungary.\\
$^{5}$MIT, Cambridge, USA.\\
$^{6}$Institute of Nuclear Physics, Cracow, Poland.\\
$^{7}$Gesellschaft f\"{u}r Schwerionenforschung (GSI), Darmstadt, Germany.\\
$^{8}$Joint Institute for Nuclear Research, Dubna, Russia.\\
$^{9}$Fachbereich Physik der Universit\"{a}t, Frankfurt, Germany.\\
$^{10}$CERN, Geneva, Switzerland.\\
$^{11}$Institute of Physics \'Swi{\,e}tokrzyska Academy, Kielce, Poland.\\
$^{12}$Fachbereich Physik der Universit\"{a}t, Marburg, Germany.\\
$^{13}$Max-Planck-Institut f\"{u}r Physik, Munich, Germany.\\
$^{14}$Institute of Particle and Nuclear Physics, Charles University, Prague, Czech Republic.\\
$^{15}$Department of Physics, Pusan National University, Pusan, Republic of Korea.\\
$^{16}$Nuclear Physics Laboratory, University of Washington, Seattle, WA, USA.\\
$^{17}$Atomic Physics Department, Sofia University St. Kliment Ohridski, Sofia, Bulgaria.\\ 
$^{18}$Institute for Nuclear Research and Nuclear Energy, Sofia, Bulgaria.\\ 
$^{19}$Institute for Nuclear Studies, Warsaw, Poland.\\
$^{20}$Institute for Experimental Physics, University of Warsaw, Warsaw, Poland.\\
$^{21}$Rudjer Boskovic Institute, Zagreb, Croatia.\\

\vspace*{3mm}

\noindent
{\bf ABSTRACT}

\vspace*{1.5mm}

\begin{abstract}

Transverse momentum spectra of $\pi^{\pm}$, $p$, $\bar{p}$, 
$K^{\pm}$, $K^0_s$ and $\Lambda$ at midrapidity were 
measured at high $p_T$ in Pb+Pb collisions at $158\,\mathrm{GeV/nucleon}$ 
beam energy by the NA49 experiment. 
Particle yield ratios ($p/\pi$, $K/\pi$ and $\Lambda/K^0_s$) show 
an enhancement of the baryon/meson ratio for $p_T>2\,\mathrm{GeV/c}$. 
The nuclear modification factor $R_{CP}$ is extracted and 
compared to RHIC measurements and pQCD calculations.

\end{abstract}

\input{proc_00.tex}

\input{proc_01.tex}

\input{proc_02.tex}

\input{proc_03.tex}

\input{proc_04_ref.tex}
\end{document}

%% file: proc_00.tex
\section{INTRODUCTION}

One of the most interesting features discovered at RHIC
is the suppression of high $p_{T}$ particle production in central 
nucleus-nucleus reactions relative to peripheral ones or p+p collisions.
This is generally interpreted as a sign of parton energy loss in hot and dense nuclear matter. 
Additionally, an enhancement of baryon/meson ratios above unity at high $p_{T}$ was observed and can be explained 
in the context of quark coalescence models. 
The aim of this analysis is to investigate the energy dependence of these effects 
by studying nucleus-nucleus reactions at top SPS energy ($\sqrt{s_{NN}}=17.3\,\mathrm{GeV/nucleon}$)
(see: \cite{Adle04,Ruan05}).

%% file: proc_01.tex
\section{DATA ANALYSIS}

Centrality selection is based on a calorimetric measurement of the 
energy observed in the projectile spectator region of phase space (see \cite{Afan99}). 
Charged particle spectra ($\pi^{\pm}$, $p$, $\bar{p}$ and $K^{\pm}$) in the center of mass rapidity interval $[-0.3,0.7]$ 
are analyzed in the centrality ranges 
(0-5)\%, (12.5-23.5)\%, (33.5-80)\% of the total inelastic cross section. 
The tracking efficiency for single tracks is above 95\% and an efficient fake 
track rejection is applied. 
The particle identification is done by unfolding the energy loss spectra measured in different phase-space bins. 
The typical $\frac{\mathrm{d}E}{\mathrm{d}x}$ resolution varies between 3 and 6\%. The 
$\pi^{\pm}$ and $p$, $\bar{p}$ yields were not corrected for feed down from the 
decay of $K^0_s$ and hyperons; furthermore the $K^{\pm}$ yields were not corrected 
for decay loss.

Neutral strange particles were analyzed in the centrality range (0-23.5)\%.
They are identified via the topology of their weak decay into the channels $K^{0}_{s} \rightarrow \pi^{+} \pi^{-}$ ($\mathrm{BR}=68.95\%$) and $\Lambda \rightarrow p \pi^{-}$ (63.9\%).
For the V0-candidates, selected by geometrical criteria, the invariant mass of the daughter particles is calculated as a function of $p_T$  and the yields of $K^0_s$ and $\Lambda$ are extracted on a statistical basis. 
The shown results are for the rapidity interval $[-0.5,0.5]$ and corrected for acceptance and reconstruction inefficiency.
The $\Lambda$ yields are not corrected for feed down from the decay of heavier hyperons.

%% file: proc_02.tex
\section{PHYSICS RESULTS}

\begin{figure}[!ht]
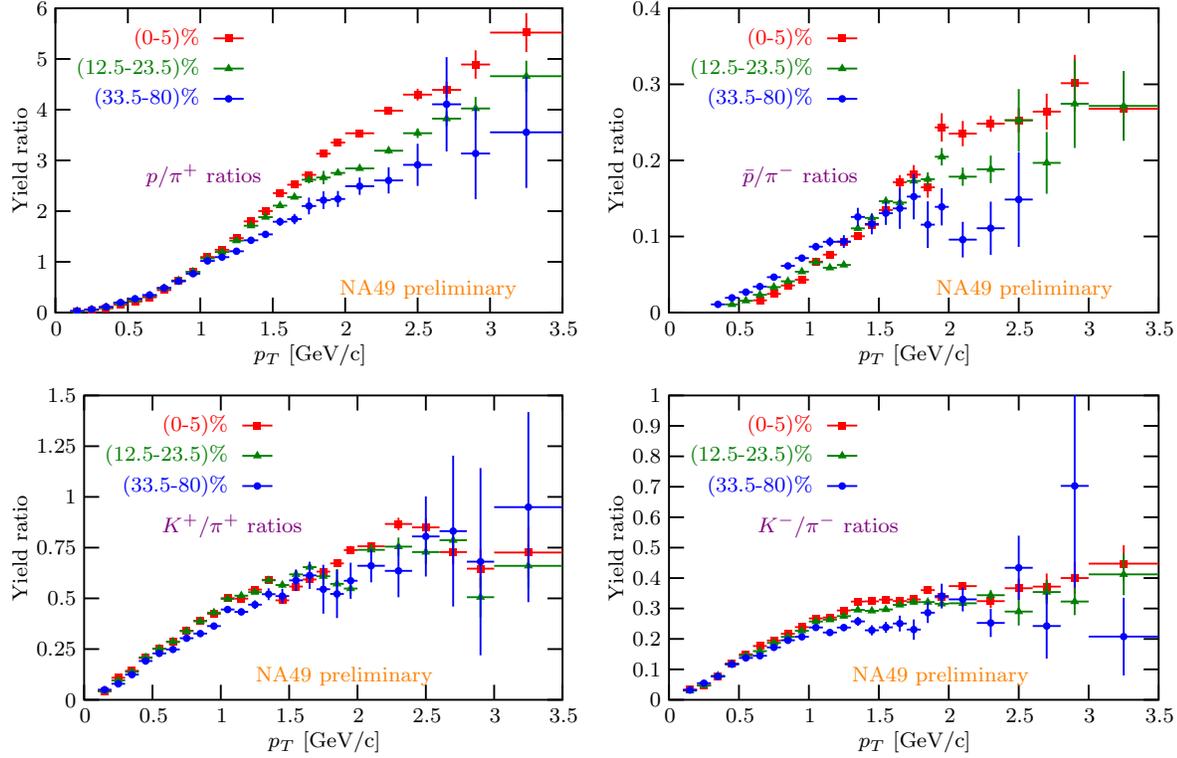

\begin{tabular}{ll}
{\scriptsize\blopeps[width=7.5cm, height=5cm]{fig/Q/Q.Pb+Pb.p.1.beps}} & {\scriptsize\blopeps[width=7.5cm, height=5cm]{fig/Q/Q.Pb+Pb.p.-1.beps}} \\
{\scriptsize\blopeps[width=7.5cm, height=5cm]{fig/Q/Q.Pb+Pb.K.1.beps}} & {\scriptsize\blopeps[width=7.5cm, height=5cm]{fig/Q/Q.Pb+Pb.K.-1.beps}} \\
\end{tabular}
\caption{\footnotesize Proton/pion (upper panels) and kaon/pion (lower panels) ratios vs.\ $p_T$ and centrality.}
\label{ratio1}
\end{figure}

The proton/pion and the kaon/pion ratios are shown in 
Fig.\ \ref{ratio1}. These ratios exhibit a monotonic increase with 
$p_T$ and centrality at high $p_T$. The kaon/pion ratios show a 
saturation tendency at high $p_T$, particularly the $K^{-}/\pi^{-}$ ratio.

\begin{figure}[!ht]
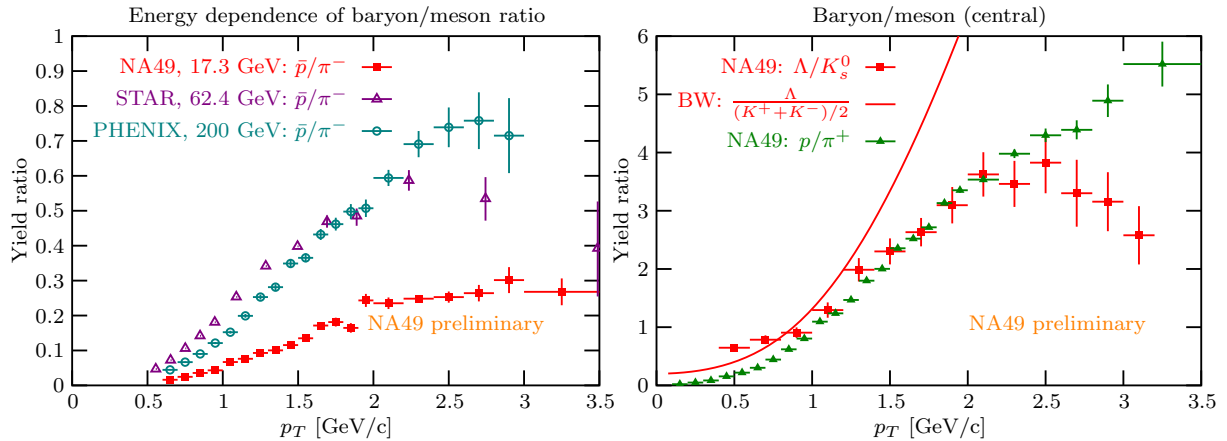

{\scriptsize\blopeps[width=8cm, height=6cm]{fig/Q/BMr2.beps}\blopeps[width=8cm, height=6cm]{fig/Q/BMr0.1.beps}}
\caption{\footnotesize The energy dependence of proton/pion ratios (left 
panel), and a comparison of the baryon/meson ratios at top SPS energy to a 
Blast-Wave parametrization (right panel).}
\label{ratio2}
\end{figure}

In the left panel of Fig.\ \ref{ratio2}, our measurement of proton/pion ratio is 
compared to RHIC data. The shape of these curves is approximately 
energy independent. The right panel of Fig.\ \ref{ratio2} shows NA49 
baryon/meson ratios, compared to a Blast-Wave (BW, see \cite{Reti04}) 
parametrization of $m_T$ spectra and radius parameters from Bose-Einstein correlations 
of pions, fitted simultaneously at low $p_T$. The BW model curve does not describe the data at 
high $p_T$.

\begin{figure}[!ht]
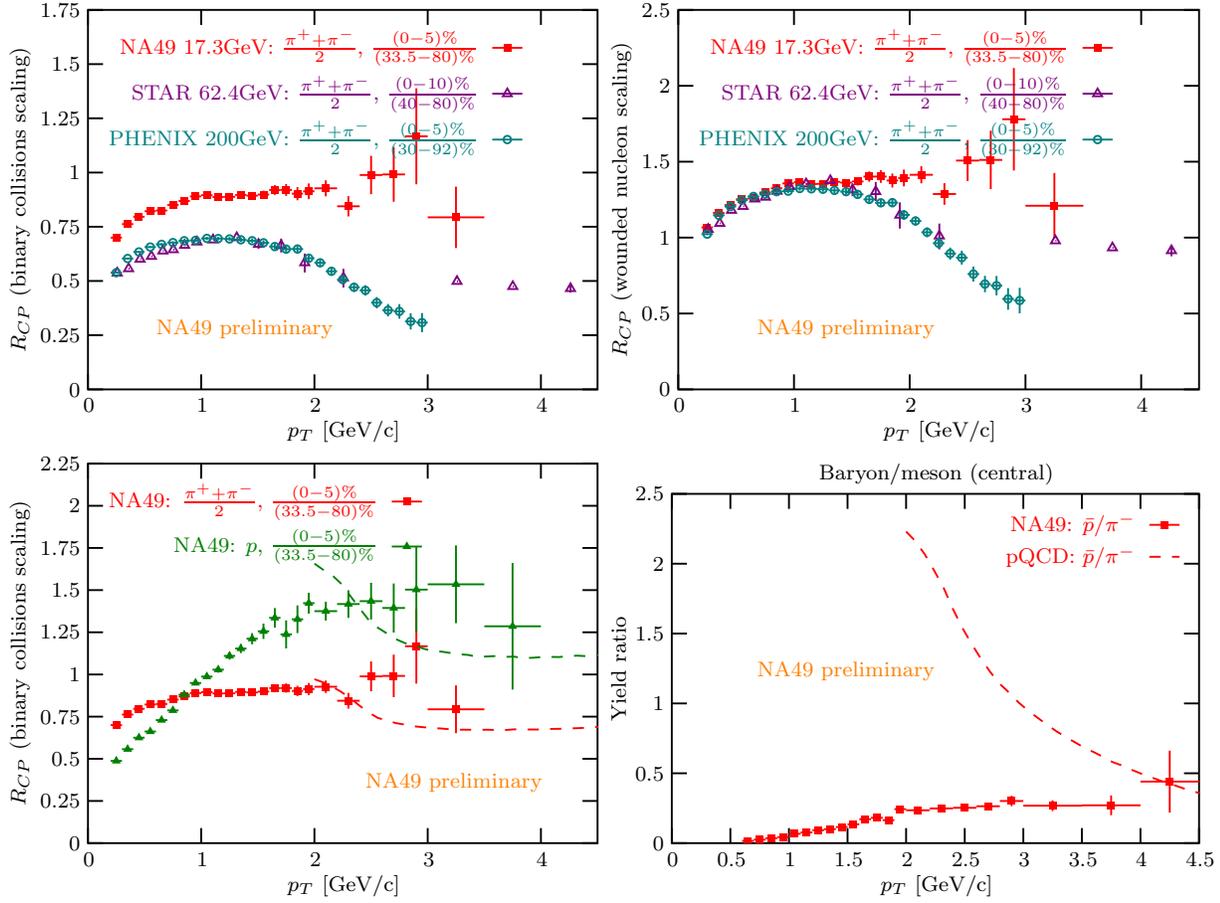

{\scriptsize\blopeps[width=8cm, height=6cm]{fig/R/R.Pb+Pb.B.beps}\blopeps[width=8cm, height=6cm]{fig/R/R.Pb+Pb.W.beps}}
{\scriptsize\blopeps[width=8cm, height=6cm]{fig/R/R.Pb+Pb.XNW.beps}\blopeps[width=8cm, height=6cm]{fig/Q/BMr.XNW.beps}}
\caption{\footnotesize Energy dependence of 
$R_{CP}$ vs.\ $p_T$ (upper panels): binary collision scaling (left panel) and 
wounded nucleon scaling (right panel). Comparison of data to 
pQCD calculations (lower panels): the nuclear modification factor $R_{CP}$ (left panel) and the 
baryon/meson ratio (right panel).}
\label{rcp}
\end{figure}

The nuclear modification factor $R_{CP}$ is defined by 
$R_{CP}:=\frac{N(\mathrm{Peripheral})}{N(\mathrm{Central})}\cdot\frac{\mathrm{Yield}(\mathrm{Central})}{\mathrm{Yield}(\mathrm{Peripheral})}$. 
Here $N$ can be either the number of binary collisions or the number of 
wounded nucleons obtained from model calculations in the given centrality 
range. 
The upper panels of Fig.\ \ref{rcp} show the energy dependence of $R_{CP}$ 
vs.\ $p_T$ of pions with binary collision and with wounded nucleon scaling.
At high $p_T$ there is a strong energy dependence with 
both scalings, however at low $p_T$ wounded nucleon scaling makes 
$R_{CP}$ energy independent. A similar phenomenon was pointed out for 
unidentified particles in \cite{Back05}.
The lower panels of 
Fig.\ \ref{rcp} show the comparison of our data to pQCD calculations 
(see \cite{Wang04}). $R_{CP}$ is consistent with the pQCD calculation 
at $p_T>2\,\mathrm{GeV/c}$. However, the pQCD prediction 
for the antibaryon/meson ratio is very far from the data below $4\,\mathrm{GeV/c}$.

%% file: proc_03.tex
\section{CONCLUDING REMARKS}

First NA49 results on particle yields around midrapidity in the range $2\,\mathrm{GeV/c}\leq p_T <4.5\,\mathrm{GeV/c}$ 
were presented from a study of $158\,\mathrm{GeV/nucleon}$ beam energy Pb+Pb collisions.

A monotonic increase of baryon/meson ratios and kaon/pion ratios with $p_T$ and centrality was observed at high $p_T$. 
The $p_T$ shape of the baryon/meson ratio is approximately energy independent. 
The measured baryon/meson ratios were compared to a Blast-Wave model: the 
model predictions exceed the data for $p_T>1.5\,\mathrm{GeV/c}$.

The nuclear modification factors $R_{CP}$ were also determined from the particle yields for various particle species, 
as a function of $p_T$. 
The measured $R_{CP}$ ratio does not show Cronin enhancement for the mesons at larger $p_T$ when using 
binary collision scaling. The behavior is qualitatively similar to the $p_T$ 
shape observed at RHIC. 
A strong energy dependence of the $R_{CP}$ ratios was observed at high $p_T$ with both binary collision and 
wounded nucleon scaling. However, at low $p_T$, the 
wounded nucleon scaling factorizes out the energy dependence. 
Results for $R_{CP}$ with binary collision scaling are consistent with 
pQCD model calculations at $p_T>2.5\,\mathrm{GeV/c}$. 
However, the pQCD calculation strongly overpredicts the observed antibaryon/meson 
ratio for $p_T<4\,\mathrm{GeV/c}$.

\vspace*{4mm}

\noindent
{\bf ACKNOWLEDGEMENTS}

\vspace*{2mm}

This work was supported by the US Department of Energy
Grant DE-FG03-97ER41020/
A000,
the Bundesministerium fur Bildung und Forschung, Germany, 
the Virtual Institute VI-146 of Helmholtz Gemeinschaft, Germany,
the Polish State Committee for Scientific Research (1 P03B 097 29, 1 PO3B 121 29,  2 P03B 04123), 
the Hungarian Scientific Research Foundation (T032648, T032293, T043514),
the Hungarian National Science Foundation, OTKA, (F034707),
the Polish-German Foundation, the Korea Research Foundation Grant (KRF-2003-070-C00015) and the Bulgarian National Science Fund (Ph-09/05).